# The mystery of energy compensation

Lewis G. Halsey

University of Roehampton, Holybourne Lane, London SW15 4JD, U.K.

## Abstract

The received wisdom on how activity affects energy expenditure is that the more activity is undertaken, the more calories will have been burned by the end of the day. Yet traditional hunter-gatherers, who lead physically hard lives, burn no more calories each day than western populations living in labour-saving environments. Indeed, there is now a wealth of data, both for humans and other animals, demonstrating that long-term lifestyle changes involving increases in exercise or other physical activities do not result in commensurate increases in daily energy expenditure (DEE). This is because humans and other animals exhibit a degree of 'energy compensation' at the organismal level, ameliorating some of the increases in DEE that would occur from the increased activity by decreasing the energy expended on other biological processes. And energy compensation can be sizable, reaching many hundreds of calories in humans. But the processes that are downregulated in the long-term to achieve energy compensation are far from clear, particularly in humans – we do not know *how* energy compensation is achieved. My review here of the literature on relevant exercise intervention studies, for both humans and other species, indicates conflict regarding the role that basal metabolic rate (BMR) or low level activity such as 'fidgeting' play, if any, particularly once changes in body composition are factored out. In situations where BMR and low-level activity are not major components of energy compensation, what then drives it? I discuss how changes in mitochondrial efficiency and changes in circadian fluctuations in BMR may contribute to our understanding of energy management. Currently unexplored, these mechanisms and others may provide important insights into the mystery of how energy compensation is achieved.

We might imagine that if we undertake a daily exercise regime burning 300 kcal each session, that this results in our daily energy expenditure (DEE) increasing by 300 kcal. Indeed, this is the received wisdom on how activity affects energy expenditure - the more activity is undertaken, the more calories will have been burned by the end of the day (F.A.O./W.H.O./U.N.U 2001; World_Health_Organisation 2014). Yet there has been recognition for many centuries, in the scientific literature at least, of 'physiological limitation' or 'material compensation' going back as far as Aristotle (Egerton 1973) and considered similarly by Darwin ('balancement of growth'; Darwin 1894) and by Rubner who argued that not all organs can be in a state of high activity simultaneously (Rubner 1910). And there is now a wealth of data, both for humans and other endothermic animals, demonstrating that lifestyle changes involving chronic increases in exercise or other physical activities do not result in commensurate increases in daily energy expenditure (DEE). Rather, humans and animals exhibit at least a degree of what herein will be called 'energy compensation' at the organismal level, ameliorating some of the increases in DEE that would occur from the increased activity by decreasing the energy expended on other biological processes (Pontzer 2017, 2021). From the perspective of the 'Principle of Allocation' upon which life history theory was founded (Sibly and Calow 1986), energy compensation is to be expected because given limited resources an animal will benefit from trade-offs in the allocation of those resources to various endogenous processes (Glazier 2009), which could include not only activity but also feeding, growth and reproduction.

Yet, at present, work from the field and the lab for both humans and other animals draws contradictory conclusions about the biological processes underpinning energy compensation. It is often unrealised that literature is very unclear on what aspects of physiology and/or behaviour are downregulated, and there may be key factors of importance that have not yet even been considered. In this review I discuss the lines of evidence that concur and conflict, and what future studies are probably necessary, to elucidate the behavioural-physiology of energy compensation. I focus on energy compensation in response to activity, but note here that energy compensation may arise in response to increases in energy for other processes such as growth (Reid et al. 2011; Sears 2005) and reproduction (Becker et al. 2013; Koch and W 1983).

**Evidence for the existence of energy compensation**

DEE adjusted for weight and age is similar between human populations of developing and industrialised nations around the world despite the diversity of lifestyles and wide range of physical activity levels represented by those groups (Dugas et al. 2011; Pontzer 2018; Pontzer et al. 2012). Analogously, while animals in captivity are considerably less active than their wild counterparts as observed in, for example, macaques (Macaca nigra; Melfi and Feistner 2002), gibbons (Hylobates lar; Warren 2010), chimpanzees (*Pan troglodytes*) and gorillas (*Gorilla gorilla*) (Ross and Shender 2016), and tigers (Panthera tigris; Breton and Barrot 2014), the energy expenditures of wild and captive animal populations are similar (Munn et al. 2013; Nie et al. 2015; Pontzer et al. 2014; Stephenson et al. 1994). The findings of intervention studies concur with these observations. Human participants exhibit a smaller increase in DEE than expected when prescribed daily exercise levels are increased (Dhurandhar et al. 2015; Garland et al. 2011; Goran and Poehlman 1992; Hand et al. 2020; Herrmann et al. 2015; Keytel et al. 2005; Willis et al. 2020; Wing 1999). Moreover, the longer and greater the exercise intervention, the greater the estimated energy compensation exhibited (for a summary see Pontzer 2018; his Figure 2). Similarly, experimental studies on animals have usually found that individuals obliged to do more physical work in order to gain a unit of food nonetheless exhibit a limited increase in daily energy expenditure (Lark et al. 2018; O'Neal et al. 2017; Pontzer 2015).

These data representing multiple species in the field and laboratory provide clear evidence that for extended periods of time during which overall activity levels are increased and thus the energy expended during that activity (activity energy expenditure; AEE) increases, there is a reduction in the energy expended by the body on certain processes that at least partially compensates. But which processes compensate is somewhat of a mystery. Thurber et al. (2019) write that the processes underlying metabolic compensation in ultra-distance runners probably include a '*reduction in nonexercised activity and reduction of physiological activity in other organ systems*', while Pontzer (2018) argues that the evidence to date for both humans and other animals suggests '*changes in other non-musculoskeletal physiological activity contribute to energy compensation*'. These broad statements in major, recent papers investigating energy compensation reflect the fact that while energy savings have been documented a multitude of times, we do not yet understand where the body makes those savings.

**The state of the art regarding how energy compensation is achieved**

Publications on this topic give credence to the idea that in order to compensate for higher energy expenditures during activity, two broad categories of the energy budget could in principle be reduced. First, there are the costs of low level activity, formerly termed 'non-exercise activity thermogenesis' (NEAT; Garland et al. 2011) or 'spontaneous physical activity' (SPA; Martin et al. 2007), which includes fidgeting (Levine et al. 1999; Mehrabian and Friedman 1986) and 'pottering' as well as postural costs such as to sit or stand (Popp et al. 2018; Tickle et al. 2012). Second, there are the costs of physiological processes that contribute to basal metabolic rate (BMR) and thus daily basal energy expenditure (BEE; an estimated value of total basal energy expenditure assuming that the measurement of BMR is a constant).

*Evidence for reductions in NEAT*

Many, perhaps most, studies and reviews of energy compensation have assumed that energy-saving changes in NEAT is the process at play (e.g. Goran and Poehlman 1992; Hand et al. 2020; Meijer et al. 1999; Morio et al. 1998), and some work seems to provide indirect evidence for this. For example, an across-school study of children reported that the amount of intense physical activity they undertook at school did not relate to their total levels of activity over the entire day, i.e. levels of NEAT were compensating for levels of intense physical activity (Mallam et al. 2003), and similarly, a study of elderly participants found they exhibited no increase in daily activity levels during periods when they participated in a physical training intervention (Meijer et al. 1999). A recent experiment on overweight women reported that exercise did not induce an increase in their daily energy expenditure nor in their resting energy expenditure, and argued that this left NEAT as the facet of energy expenditure that was decreasing to compensate (Riou et al. 2019).

However, NEAT has also been reported unchanged (e.g. Blaak et al. 1992; Rangan et al. 2011; Willis et al. 2020) or increasing (e.g. Hollowell et al. 2009; Meijer et al. 1991; Westerterp et al. 1992) in response to heightened levels of exercise, and anyway most studies have not directly measured it (Melanson 2017). Reviews of the literature to determine whether NEAT in humans decreases to compensate or partially compensate for increases in activity energy expenditure deem the evidence to be conflicting (Fedewa et al. 2017; Melanson 2017; Melanson et al. 2013; Washburn et al. 2014). The main conclusion stated by all four papers is that there is not the evidence overall in the literature to infer that NEAT systematically decreases in response to either short-term or long-term increases in daily exercise levels. I summarise their pertinent, subsidiary conclusions here: i) The energy costs of NEAT are a strong predictor of DEE and vary widely within and between people. ii) This variation may in part be explained by marked individual differences in NEAT-driven compensatory responses and the fact that few studies thus far have measured NEAT directly. iii) Shorter exercise sessions may influence NEAT less than do longer sessions, while decreases in NEAT may attenuate through the exercise *period*.

The animal literature here is dominated by mouse studies. These papers arguably provide clearer evidence of reductions in NEAT in response to greater activity levels (wheel running) than do the human studies (De Carvalho et al. 2016; O'Neal et al. 2017). However, the resulting degree of compensation appears fairly small (~5% reduction in DEE represeting an attenuation in the increase in DEE due to wheel running of about 20%; Lark et al. 2018), which may mean that other energy compensation mechanisms are also at play. And these could include the arguable confound of reducing non-shivering thermogenesis in response to the muscle thermogenesis from wheel running (Even and Blais 2016), because mice might often be housed at ambient temperature below their thermoneutral zone (O'Neal et al. 2017; Speakman and Keijer 2013). A study of responses by starlings *Sturnus vulgaris* to food insecurity offers inconsistent evidence for reduced physical activity (Bateson et al. 2021). For human studies at least, more sophisticated investigations are required to clarify the relationship between activity energy expenditure and NEAT, employing methods that can study NEAT directly (Dugas et al. 2011), perhaps, for example, with accurately calibrated, sensitive activity monitors.

*Evidence for reductions in BMR*

The theory behind the idea that BMR (and thus BEE) decreases and therefore compensates for increases in AEE is that certain physiological processes required to maintain homeostasis such as perhaps immune-competency, protein turnover and somatic repair are somehow de-prioritised and thus become down-regulated when energy compensation is a more pressing homeostatic driver (Pontzer 2018; Wiersma and Verhulst 2005). (NB, whether such changes in metabolism are driven by top-down regulatory mechanisms or are the result of bottom-up inter-cellular competition for resources needs further debate, elsewhere; see e.g. Archer et al. 2018b). Such downregulation could explain increases in oxidative stress and DNA damage in animals required to apply more 'effort' (for a review, see Soulsbury and Halsey 2018). *In extremis*, this downregulation could slow growth or cause the onset of disease and impairments to ovulation and reproduction (Lebenstedt et al. 1999; Melin et al. 2015; Perrigo and Bronson 1983). However, the actual contribution of changes in BMR to energy compensate for increases in AEE is far from clear (Herrmann et al. 2015).

It is widely accepted by most, but not all (Mitchell et al. 2017), that BMR decreases in response to a negative energy balance at the organismal level (see also Martins et al. 2020), even when statistically accounting for changes in body mass and condition, both in humans (Hopkins et al. 2014; Johannsen et al. 2012; Leibel et al. 1995; Martin et al. 2007; Schwartz and Doucet 2010), other primates (Yamada et al. 2013), and rodents (Hambly and Speakman 2005). Interventions that increase activity levels often lead to changes in body composition such as weight loss and increased levels of fat free mass, which in turn affect BMR (Silva et al. 2012), making it difficult to justify associating energy compensation due to a decrease in BMR directly with increased activity. To be clear – in response to exercise interventions that induce weight loss, even if a decrease in BMR is adjusted for changes in body composition such as overall weight, fat free mass and/or fat mass, (which may be very difficult; Heymsfield et al. 2018), we cannot assert that this downregulation in metabolic rate is not at least in part the body responding to negative energy balance (Tremblay et al. 2013). Only studies that report a decrease in BMR alongside increases in activity levels and no substantive decrease in body mass can reasonably be used to infer that energy compensation in response to heightened activity *per se* can include downregulation of BMR.

It is important to recognise that, in the short term, BMR often increases in response to increases in activity. EPOC – excess post-exercise oxygen consumption – is the elevation of metabolic rate above resting levels for a period after the completion of exercise, which tapers away over time. Extensive literature on aerobic and resistance exercise, reviewed by Laforgia et al. (2006) and Farianatti et al. (2013), indicate that EPOC typically lasts up to a few hours. Longer bouts of activity, however, can result in EPOC continuing for many hours (Melby et al. 1993; Short and Sedlock 1997; Tuominen et al. 1996), sometimes a day (Bielinski et al. 1985; Maehlum et al. 1986) or even two days (Sjödin et al. 1996; Williamson and Kirwan 1997). Therefore, only studies that compare BMR against activity levels for the long term have the potential to record a decrease in BMR (and thus BEE) (Pontzer 2018) (Figure 1). Figure 1 also serves to highlight that AEE is often calculated rather than measured; in these cases it is determined by subtracting BEE (and sometimes also digestion costs) from DEE and thus inaccuracies in the former will result in over- or understimations in AEE.

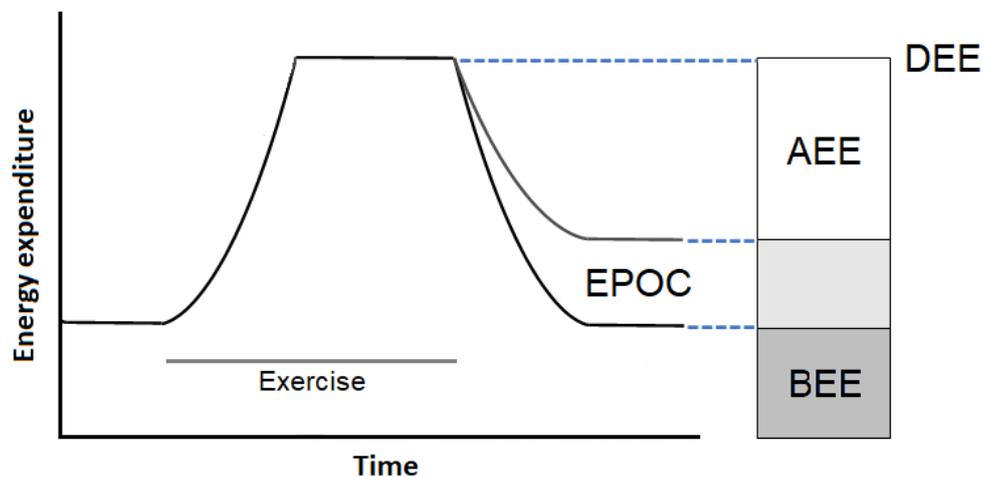

Figure 1. The effects of excess post-exercise oxygen consumption (EPOC) on the disaggregation of daily energy expenditure (DEE) into basal energy expenditure (BEE) and activity energy expenditure (AEE). If EPOC subsequent to activity is still present at the point that an attempt is made to measure basal metabolic rate (BMR), BEE will be overestimated and in turn AEE will be underestimated.

I find very limited evidence in the human literature of BMR decreasing in response to a chronic period of increased activity energy expenditure *per se*. In most studies where participants exhibited no change or only a marginal decrease in body mass in response to an increase in daily activity levels they also exhibited no change in BMR (often calculated independent of body condition) (Colley 2008; Colley et al. 2010; Flack et al. 2020; Goran and Poehlman 1992; Hand et al. 2020; Herrmann et al. 2015; Meijer et al. 1999; Riou et al. 2019; Van Etten et al. 1997; Willis et al. 2014). In a few studies where participants experienced no change in body mass they exhibited a small increase in BMR (Hunter et al. 2000; Morio et al. 1998; Withers et al. 1998). Two papers report a decrease in BMR despite little to no change in body weight (Silva et al. 2017; Westerterp et al. 1992), but the decrease is modest and would only account for a small proportion of the observed energy compensation in those studies (Pontzer 2015). A meta-analysis by MacKenzie-Shalders et al. (2020), which focussed on studies of healthy non-elderly participants, though typically the participants were overweight, found that papers reporting a stable body mass during the experimental intervention also reported an increase in resting metabolic rate (NB, none of these studies included a dietary component to the intervention). While they argue that in some studies resting metabolic rate might be overestimated due to EPOC if the measurement was taken too soon after the most recent bout of the exercise intervention (Figure 1), nonetheless there is no evidence from this work of a decrease in BMR in response to an increase in AEE.

On the other hand, a reanalysis of the Westerterp (1992) study by Careau (2017) suggests that when this repeated-measures dataset is analysed for not only across-individuals but within-individual correlations, there is clear evidence of energy compensation within individuals based on the slope of the relationship between DEE and BEE being <1. Caution, however, must be taken with such analyses because the statistical effect of regression dilution will to some degree underestimate the true slope value creating artefactual energy compensation (Halsey and Perna 2019). Colleagues and I recently conducted an analysis on another repeated-measures dataset, comprising paired energy expenditure data for a field survey of 1756 elderly individuals, the first set of measurements taken between 1998 and 2000, and the second set about 7 years later. Within-individuals, variations in AEE were negatively associated with variations in BMR and thus BEE accounting for age, fat free mass and fat mass (Careau and al. Submitted), and subsequent analysis confirms that AEE and body mass are not correlated (Vincent Careau, pers. comm.). These findings, then, are the first to provide strong evidence, being based on a large sample size of direct metabolic rate measurements, of substantial decreases in BEE resulting from increases in AEE *per se*. Specifically, the data suggest that the decreases in BMR can compensate for more than a quarter of the increase in energy expenditure due to increases in activity (Careau and al. Submitted).

*Does the animal literature mirror the human literature on BMR-driven compensation?*

Microchiroptera bats reduce their resting metabolic rate after a period of flying (Speakman and Racey 1991), but I am not aware of other field studies that have reported, or investigated, this phenomenon. There are, however, a number of lab-based exercise intervention studies on birds and mice that measured BMR and DEE. In 2015, Herman Pontzer reanalysed these studies to show that, in most cases for both birds and mammals, when activity levels increased (such as more time spent flying or wheel-running) basal metabolic rate decreased, sometimes subtly and sometimes substantially (Pontzer 2015; his Figure 2). However, in all of the studies reviewed by Pontzer (2015) that visually demonstrate a substantial decrease in BMR, daily food intake and body mass had also decreased substantially in association with the increased activity levels (Bautista et al. 1998; Perrigo and Bronson 1983; Tiebout 1991; Vaanholt et al. 2007). In contrast, for those studies that visually indicate at most only a subtle decrease in BMR (Deerenberg et al. 1998; Weimerskirch et al. 1992; Westerterp et al. 1992; Wiersma and Verhulst 2005), body mass also decreased at most only slightly. There is one exception: in a study of starlings, body mass decreased quite a lot and yet BMR decreased only modestly (Wiersma et al. 2005). This is also the one study where food intake increased as activity energy expenditure increased. Overall then, interpretation of exercise interventions in the animal literature across birds and mammals concurs with that for the human literature – reductions in BMR due to increased activity energy expenditure *per se* are modest at best when body mass remains constant.

Field data encourage a contra conclusion, albeit with caveats. Reduction in basal metabolism appears to have been observed, indirectly, in a range of animal species in the field through long-term measurements of heart rate. Heart rate correlates with metabolic rate in endothermic species (Green 2011). The harder an animal is working the faster its heart beats to support upregulation of the cardio-respiratory system, in accordance with Fick's principle (Fick 1870). The lowest mean heart rate per day probably represents heart rate while the body's metabolism is basal, and changes in daily lowest mean heart rate probably indicate changes in daily basal metabolic rate. Halsey et al.

(2019) found that individuals of many species including birds, mammals and fish exhibited lower minimum heart rates during periods when their daily heart rate, and thus daily metabolic rate, was higher. This suggests that during periods when activity energy expenditure is higher, basal metabolic rate decreases. Indeed, investigating the year-round data sets that were available for red deer *Cervus elaphus*, alpine ibex *Capra ibex* and greylag geese *Anser anser* showed that during months when daily heart rate was higher (which tended to coincide with key annual events such as reproduction) the slope between daily heart rate and minimum heart rate was particularly shallow indicating strong energy compensation through the decrease of basal metabolic rate. There is not, however, information on body mass changes over time and so we cannot rule out the possibility that reductions in BMR are driven by body mass loss rather than heightened AEE *per se*. There is also the possibility that during periods when activity levels are high, the heart adaptively increases in size, affecting the relationship between heart rate and metabolic rate; a slower, larger heart could be associated with the same metabolic rate as a faster, smaller heart. Thus the aforementioned relationships for deer, ibex and geese are conceivably driven by changing levels of physical fitness.

A dataset of more direct measures of metabolic rate alongside changes in body mass is necessary to confidently ascertain whether heightened AEE *per se* is associated with a downregulation of BMR in animals. Until then, it is worth noting that the findings from the aforementioned heart rate-based data for animals in the field (Halsey et al. 2019) are analogous to those of the study mentioned earlier on humans in the field (Careau and al. Submitted), perhaps offering support that the animal data can indeed be interpreted at face value.

*Summary of the literature on BMR- and NEAT-driven compensation*

Synthesising the findings of the various papers cited above does not return a clear conclusion about the importance of BMR in energy compensation in response to AEE increases. The results of a field survey for humans and field surveys for other animals, where variations in activity energy expenditure are not usually determined by prescribed exercise regimes, suggest that BMR downregulation can be a substantial, perhaps the substantial, element of energy compensation. In contrast, lab-based exercise interventions indicate a limited role, if any, of BMR. And for humans specifically, lab-based studies also indicate an unclear role for NEAT. When a reduction in NEAT is recorded it can rarely account for a substantial proportion of the observed energy compensation (Colley et al. 2005; Lark et al. 2018; Pontzer 2015; Pontzer et al. 2016). In contrast, the few experiments of NEAT in animals, all on mice, provide clearer evidence of its substantive role in energy compensation.

*Investigating three studies in detail*

What then explains the substantial energy savings reported in many human studies? Certain published papers are the most suitable, based on their data sets, for exploring the possible contributions of both NEAT and BMR to observed energy compensation.

First, Thurber et al. (2019) measured the daily energy expenditure, BMR and body condition of six athletes participating in a transcontinental marathon event called the Race Across the USA (RAUSA). While the runners had very high daily energy expenditures due to running a marathon each day, they also energy-compensated for the running AEE (~3000 kcal) by an average of 600 kcal/d, according to calculations of predicted versus measured DEE. None of this energy compensation is explained by decreases in absolute BMR, which if anything marginally increased (despite a slow but cumulatively substantive decrease in body mass of 4.1 kg over ~18 weeks). Potentially, a considerable proportion of this 600 kcal can be explained by a reduction in NEAT or fidgeting-like behaviours (Levine et al. 1999; Ravussin et al. 1986), for example sitting or lying motionless rather than sitting or standing while fidgeting (Levine et al. 2000), perhaps driven by fatigue. While NEAT was not measured in this study, if we assume the runners were sleeping for 8 h/d and running for 5h/d (leaving 11 h/d to vary the amount of fidgeting-like behaviours), based on measured energy costs of fidgeting in Levine et al. (2000, their Table 1), by mostly resting without fidgeting a runner might save up to 700 kcal/d compared to their typical fidgeting-related energy expenditure during periods outside of ultra-distance events. This is very similar to the maximum energy expenditure due to increases in fidgeting-like behaviours by participants in response to overeating reported by Levine et al. (2000), though this value might be considered high because 700 kcal is the typical energy expended to run about 10 km. An alternate possibility is that the metabolic rate of various tissues decreased during the long periods of running because so much available energy was being diverted to the skeletal muscle – an example of the argument that different tissues compete, often asymmetrically, for energy (Archer et al. 2018b). However, taking a BMR of 1500 kcal/d in a chronically fed state and assuming this represents the metabolic costs of organs (Müller et al. 2013), even if most of the organs except the musculature and the cardiorespiratory system consumed zero energy, this would result in whole-body energy compensation of only about 300 kcal during the periods of running. Thus periodically energy-deprived organs could explain part, but not all, of the whole-body energy compensation recorded in the RAUSA athletes.

Second, Westerterp et al. (1992) published analyses for a study somewhat analogous to that by Thurber et al. (2019), of non-athletes spending 44 weeks training for their first half-marathon. The data ultimately suggest that, again, reductions in NEAT are more substantial than reductions in BMR. While reanalysis of the Westerterp et al. (1992) data, in contrast to the findings of Thurber et al. (2019), provides strong statistical evidence ($p = 0.003$) that the participants exhibited a decrease in BMR, this decrease was fairly moderate in magnitude (median sleeping metabolic rates; week 0: 6.5 MJ/d; week 40: 6.0 MJ/d) (Westerterp et al. 1992; their Table 2). During that period, body mass decreased by 2 kg but fat-free mass increased by 3 kg (Westerterp et al. 1992; their Table 4) suggesting that changes in body composition are unlikely to explain these BMR changes either through a change in body condition or because of negative energy balance at the organismal level (mean rate of mass loss: 50 g/week). This apparent change in BMR represents an energy saving per day of 120 kcal, i.e. enough to compensate for moderate activity lasting ~30 min or heavy activity lasting ~10 min. But while this downregulation in BMR is arguably of functional significance, it explains only a fraction (~10%) of the estimated energy savings of around 1200 kcal $d^{-1}$ (5 MJ; Pontzer 2015; their figure 2A,B), leaving the rest to potentially be explained by decreases in NEAT. However, surely this deficit is too large to be filled by less fidgeting and changes in posture (Figure 2).

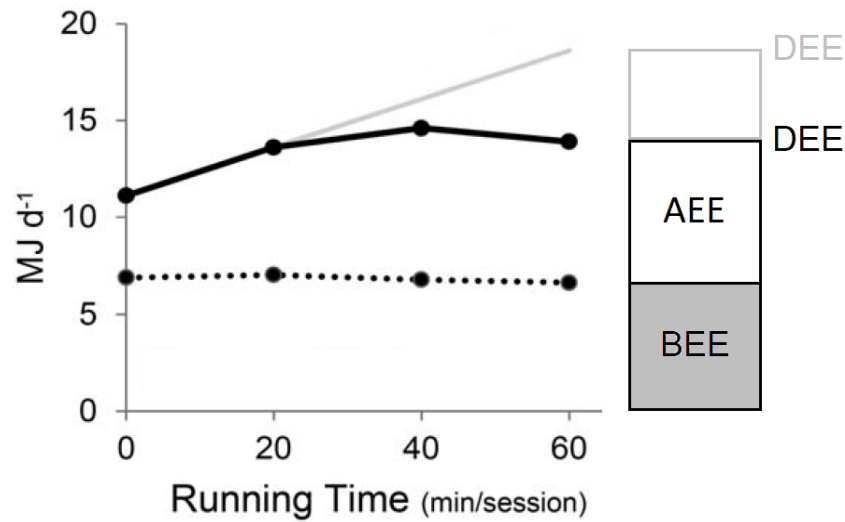

Figure 2. Daily energy expenditure, activity energy expenditure (AEE) and basal energy expenditure (BEE) of adult males training for a half-marathon over 44 weeks, during which the length of the training runs progressively increased (Westerterp et al. 1992). The grey line depicts predicted daily energy expenditure if no energy compensation occurred. BEE decreases only slightly, while the estimated decrease in AEE is considerable. Reproduced and adapted from Pontzer (2017).

Finally, data from a study that attempted to directly measure NEAT in a group exhibiting energy compensation do not indicate that these people are reducing NEAT, which thus appears to contrast the findings of Thurber et al. (2019) and Westerterp et al. (1992). However, I will argue that reduced NEAT is still a possible explanation. Daily energy expenditures of children from the UK/US and children of Shuar forager-horticulturalists are very similar despite the Shuar children being 25% more physically active and also having a greater resting energy expenditure (Urlacher et al. 2019; their Fig. 1B) (probably due to an upregulated immune system; Urlacher et al. 2018; Wolowczuk et al. 2008). Urlacher et al. (2019) report that the Shuar children seem to compensate for this by savings in their energy expenditure associated with activity, but concede that how these savings are made is unclear. Scope for humans to become more energy efficient at a given activity appears very limited. Burgess and Lambert (2010) report that the evidence for increased running efficiency in response to exercise training is mixed and, when present, modest at 3% energy savings, though savings may be greater at low speeds (Tremblay et al. (1997) report a 12, 7 and 3% energy saving when walking at 4.5, 5.5. and 6.5 km h$^{-1}$, respectively, after a 93-d training programme). This seems to leave the possibility that, similarly to the proposed mechanism for the ultra-marathon runners and the half-marathoners in training, the Shuar children reduce their fidgeting-like behaviours. Although the activity counts on the accelerometers worn on the hip by the Shuar children indicate they are considerably more active than UK and US children in total across the waking day, just possibly the Shuar children nonetheless exhibit reduced fidgeting behaviours – a difference not sensed by the hip-instrumented device due to its limited capacity to recognise NEAT (Kozey-Keadle et al. 2011) such as arm movements under certain circumstances (Fernández-Verdejo et al. 2021). Returning again to the values provided by Levine et al. (2000; their Table 1) for adults, and halving them to account for the equivalent energy expenditures of 8-year old children, in the possible scenario that children sit for half the waking day and stand for the other half but while doing so Shuar children are motionless whereas UK/US children are fidgeting, the Shuar would save 400 kcal per day compared to the UK/US children and yet the accelerometer records little of the fidgeting activity in the UK/US group. In reality of course, both groups are moderately or vigorously active for several hours per day during which we can expect the accelerometer count to accurately reflect activity levels; nonetheless there are a number of hours when they are sedentary (Urlacher et al. 2019; their Table 1) such that reduced fidgeting could provide the Shuar with substantive energy compensation.

Overall, then, these three studies considered in detail together suggest that BMR is at best a minor element of energy compensation, while NEAT is a more promising explanation but is yet to be adequately measured.

**What other aspects of energy expenditure could be involved in energy compensation?**

Studies to date tend to assume that energy compensation is explained by one or both of BMR and NEAT, but there are other possible explanations yet to be considered.

*Attenuation in daily fluctuations of BMR*

Circadian fluctuations in BMR can reach 10% (van Moorsel et al. 2016; Zitting et al. 2018) but are theorised to be attenuated in individuals that are energetically stressed, that is, in individuals that are ingesting fewer calories than required for the body to expend on all energetic processes optimally (Urlacher et al. 2019). BMR measurements are typically taken in the early morning, most commonly around 5 am, when the circadian rhythm of BMR is at its nadir. Such a measure of BMR therefore underestimates daily BEE in individuals not energetically stressed. Because AEE is typically estimated as DEE minus BEE, in turn AEE is overestimated (Figure 3A). In energetically stressed individuals, the underestimate in BEE is less and in turn the overestimate in AEE is less. When assessing the changes in BEE and AEE to energy compensate in response to increased activity levels, if those increased activity levels cause energy stress that was not previously present then any real decrease in BEE will be underestimated while any decrease in AEE will be overestimated (Figure 3B). Thus, potentially BEE plays a clearer, greater role in energy compensation than is evidenced by many studies, while NEAT or other aspects of AEE may play a smaller role. Measuring BMR closer to midnight or multiple times during the day will somewhat alleviate this problem (Figure 3).

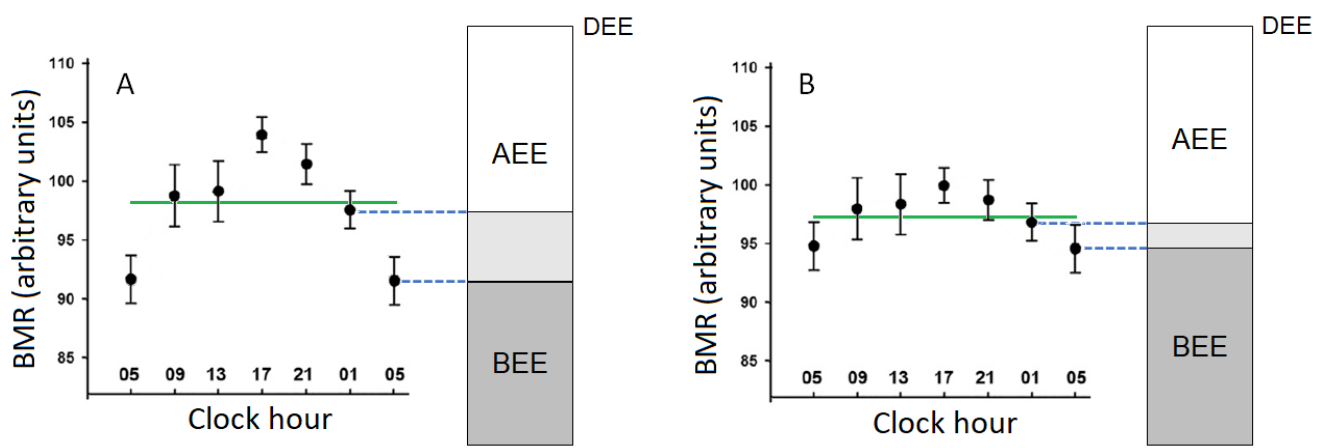

Figure 3. Calculations of basal energy expenditure (BEE; typically the total energy expended on basal processes over 24 h) and thus activity energy expenditure (AEE) depend on the magnitude of the circadian rhythm of basal metabolic rate (BMR) and the point in the circadian phase when it is measured. A) Fluctuations in BMR are substantial in individuals that are not energetically stressed such that measured BMR can be different depending on the time of day it is recorded. If recorded at a mid-point, perhaps 1 am (blue line), the measurement probably accurately estimates daily BEE (green line). If recorded at the nadir, which might typically be at around 5 am, the BMR measurement underestimates daily BEE, and in turn AEE is overestimated. B) BMR fluctuations may be attenuated in energetically stressed individuals such that BMR measured at 5 am is less of an underestimate of daily BEE and therefore the estimate of AEE is less of an overestimate. The data presented in panel A reflect laboratory measurements reported in Zitting et al. (2018; their Fig. 2); those in panel B are hypothetical.

*Changing mitochondrial efficiency*

The classical approaches to measuring energy expenditure involve recording respiratory gas exchange of the entire organism, however this represents a derivation of respiration at the sub-cellular level (Koch et al. 2021), where the relationship between oxygen consumption and energy can vary. Mitochondria are essential organelles generating the majority of energy required for cellular and physiological processes. This energy is provided in the form of ATP generated through the consumption of $O_2$ in a complex process termed oxidative phosphorylation (OXPHOS) (Koch et al. 2021). The ability of mitochondria to generate ATP, in terms of the number of ATP molecules generated for each $O_2$ atom consumed, is known as the phosphate-to-oxygen (P-O) ratio and is a function of mitochondrial 'efficiency'. (Mitochondrial efficiency is affected by the metabolic substrate being oxidised (efficiency is 15% greater when fat rather than more efficient when utilising fat compared to carbohydrates are being utilised; Welch et al. 2007). Moreover, not all $O_2$ consumption is coupled to mitochondrial ATP production, and thus some of the available free energy is subjected to proton leakage and lost as heat. This proton leakage presents a significant source of uncoupling and has been estimated to account for 20–25% of the in vivo BMR (Rolfe and Brand 1996). Scenarios where mitochondria exhibit higher efficiency of converting metabolic substrates into energy are met with decreased proton leakage and a lower metabolic rate (Murphy 2009; Stier et al. 2014). In contrast, a lower efficiency in the conversion of metabolic substrates into energy leads to higher proton leakage, heat generation and increases in metabolic rate (Salin et al. 2015). Importantly, mitochondrial efficiency can vary, both within and between individuals (Salin et al. 2015) and under different conditions. Thus, variability in mitochondrial efficiency can be a significant factor contributing to alterations in BEE (Larsen et al. 2011) and other elements of energy expenditure. Of particular significance here is the observations that both humans and animals show a mitochondrial plasticity to physical exercise (Porter et al. 2015; Stier et al. 2019). In humans, both endurance and resistance exercise promotes increased mitochondrial performance (Fernström et al. 2004; Porter et al. 2015), and there is tentative evidence that older humans with a more sedentary lifestyle have a reduced walking capacity and speed relating to a lower mitochondrial efficiency (Coen et al. 2013; Distefano et al. 2018). Changes in mitochondrial efficiency affect the energy expended on any or all energetic processes of the body (Salin et al. 2015), and hence could be the process underlying energy compensation in response to increased exercise.

**Conclusions**

The assumption that the more activity is undertaken, the more calories will have been burned by the end of the day is formalised in the 'additive' model of energy expenditure, which has been applied to both humans and other animals (Halsey et al. 2019), and underscores models of public health (F.A.O./W.H.O./U.N.U 2001; Pontzer 2015). However, clear evidence of energy compensation at the organismal level by humans and other animals has now accumulated, and it can be considerable, at least up to 600 kcal/d in humans (Thurber et al. 2019). Despite this, there are many inconsistencies and contradictions across the literature meaning that we are not in a position to state with any certainty the predominant processes involved in this compensation. Most experimental studies of humans indicate that any BMR decreases in response to increases in AEE are at best only slight, however a recent large-sample field study suggests that in older individuals it explains most of the observed energy compensation (Careau and al. Submitted). NEAT may decrease but may simply be a minor factor resulting from fatigue due to the heightened physical activity, or instead it could be the main process explaining energy compensation. In young or gestating individuals, the additional energy expenditures of growth and reproduction might be reduced (Urlacher et al. 2019). Economical energy expenditure during activity, through reduced extraneous limb movement (Fernández-Verdejo et al. 2021), more efficient biomechanics or utilising fat rather than sugar as the metabolic substrate, might possibly play a role in energy compensation in certain situations (Amati et al. 2008; Burgess and Lambert 2010; Halsey et al. 2017; Lark et al. 2018), though with the caveats that (i) biomechanical efficiency could at best only explain a fraction of substantial energy compensation and (ii) changes in metabolic substrate have only a fairly small effect on measures of metabolic rate and do not in themselves reflect changes in ATP use (Salin et al. 2015; Figure 4 A and B). For species outside their thermoneutral zone, thermoregulatory costs might be minimised because of the heat produced as a by-product of increased activity (Even and Blais 2016; O'Neal et al. 2017). Even decreases in the costs of digestion and assimilation – the thermal effect of feeding – might be part of the compensation strategy. Although digestion costs are unlikely to be a predominant factor (Morio et al. 1998) because they appear resistant to adaptation (Ocobock 2020), they can vary substantially depending on macronutrient content and thus could be affected by a change in diet (Westerterp 2004). Finally, another possible mechanism serving compensation concerns stress. Given that stress responses are attenuated in people who regularly exercise (Silverman and Deuster 2014), if indeed cortisol and epinephrine release, in response to stress, increase metabolic rate (Holland-Fischer et al. 2009; Hollstein et al. 2020), then increased activity energy costs may be compensated by decreased stress energy costs.

Another possibility is that, rather than energy compensation resulting from the downregulation of one or more supposedly physiologically discrete processes, ranging from basal costs to physical activity, it results instead from the adjustment of one fundamental driver that underpins the endogenous energy expenditure of all these processes – mitochondrial P-O efficiency and levels of proton leakage. In principle, basal processes, NEAT and so forth can all be fully maintained and yet the substrate energy used for these processes decreased by increasing P-O efficiency (Salin et al. 2015) (Figure 4). Thus ATP use is maintained while metabolic rate decreases. Moreover, rather than assuming that decreases in metabolic processes such as BMR or NEAT are driven by 'top-down' mechanisms enacted by some form of control centre, a notion I would argue is often implied in the literature including in my own offerings (Halsey 2018; Hambly and Speakman 2005; Thurber et al. 2019), we should consider viewing processes of energy compensation in terms of collaboration and competition between cells for finite energy resources (Archer et al. 2018a; Archer et al. 2018b). With this perspective, for example, a putative decrease in BMR is *caused* by a shift in the competition for energy between cells. For instance, and as discussed earlier with regards the Thurber et al. (2019) case study, at least during activity skeletal myocytes may outcompete other tissues in the acquisition and storage of consumed energy and in turn other tissues have less metabolic substrates to use resulting in their downregulation (Edward Archer, pers. comm.). Adjusting energy consumption and thus energy availability may be a way to interrogate this proposition.

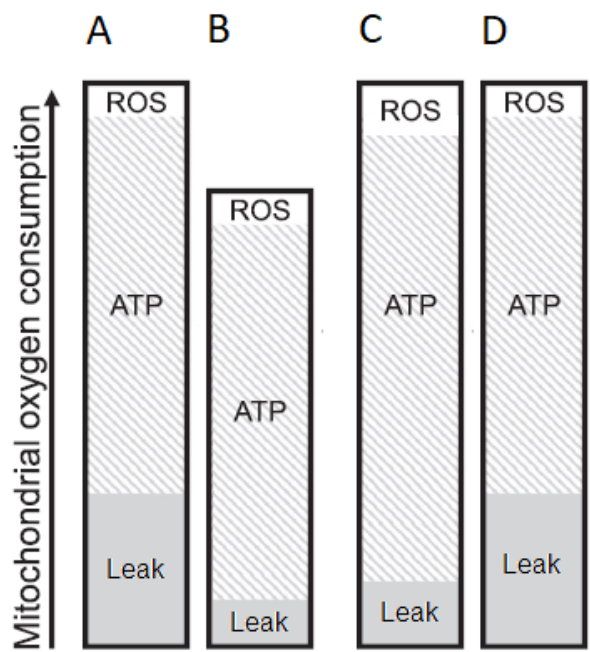

Figure 4. Comparisons of measurements of (mitochondrial) oxygen consumption between states (such as different individuals or different behavioural states within an individual) do not provide accurate comparisons of energy expenditure (quantified as ATP production) if mitochondrial 'efficiency' differs between those states. A and B: If mitochondrial efficiency is higher in state B than in A then energy use as ATP production can be the same while oxygen consumption decreases. C and D: Oxygen consumption is the same in states C and D and yet mitochondrial efficiency is lower in state D, such that energy expenditure (ATP production) is also lower. ROS = reactive oxygen species; Leak = proton leak. Adapted from Koch et al. (2021; their Figure 3).

Alongside understanding how energy compensation is achieved, there are many other important and interesting questions to investigate (Melanson 2017) including if and how the ceiling to DEE at a given time varies with factors such as body mass and age, and how activity type and intensity influence the degree and nature of energy compensation (Riou et al. 2015). Future studies need to be long-term and equipped to accurately and directly measure a full gamut of energetic processes including BMR at multiple daily timepoints, heavy physical activity and NEAT, in participants exhibiting minimal changes in body condition during the period of the exercise intervention. Perhaps, for example, this approach will elucidate a 'sweet spot' in terms of exercise volume for weight or fat loss, which trades-off increasing prescribed exercise volume to increase daily endogenous energy expenditure against the diminishing returns at greater exercise volumes due to energy compensation (Figure 5). Such studies will probably require metabolic chambers combined with sensitive physical activity monitors. It will also require the capacity to analyse the mitochondrial efficiency of samples preferably from a range of body tissues. Finally, solving the mystery of energy compensation may require recognition of substantive phenotypic variation in behavioural and physiological responses (Careau and Garland Jr. 2012) to changes in activity levels (Melanson et al. 2013), as is apparent in for example responses of hunger and circulating acylated ghrelin concentrations (King et al. 2017) and energy intake (King et al. 2008). Some people may fidget less, some people may show lower fluctuations in their BMR and some people may exhibit increased mitochondrial efficiency; others may not respond at all.

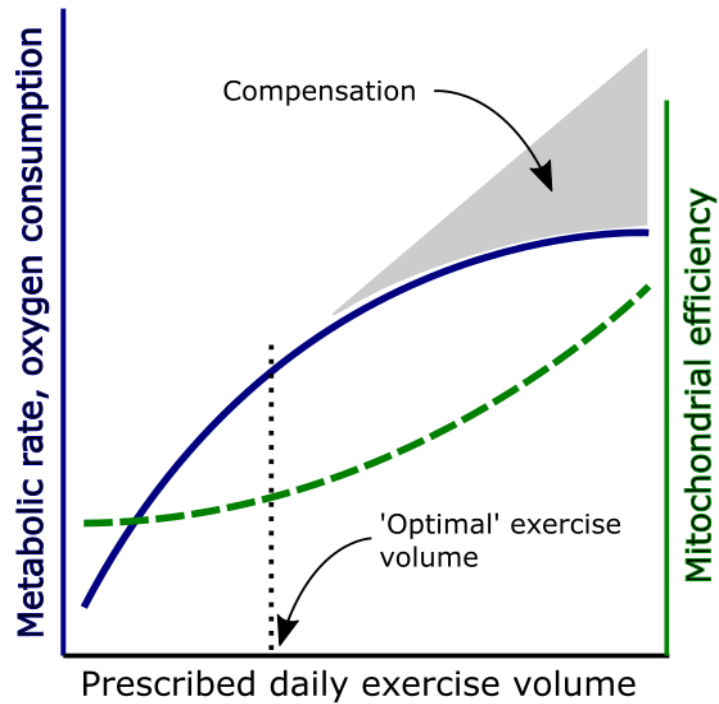

Figure 5. The relationship between prescribed daily exercise volume and the additional daily energy expenditure as a result of that exercise probably approximates a positive curve of diminishing returns, due to progressive increases in energy compensation (blue line) (see Pontzer 2018). If this energy compensation is due to increases in mitochondrial efficiency (green, dashed line), then this is energy compensation in terms of *metabolic substrate consumption* (i.e. metabolic rate), and not in terms of ATP use. At a certain volume of daily exercise, consequent increases in metabolic substrate consumption have diminished sufficiently that further increases in exercise volume offer little return – this exercise volume might be considered 'optimal' for encouraging metabolic substrate consumption.

**Acknowledgements**

Thank you to Jeff Yap for his thoughts on measures of mitochondrial efficiency, to Karine Salin for her thoughts on mitochondrial efficiency and also her ideas that form Figure 4, to Herman Pontzer for discussion about visualising the concept of a flattened BMR through the day, and to Edward Archer for discussion about how to conceive changing energy usage by the body. Jon Green gave detailed feedback on an earlier version of this article.

**Funding**

None.